  \documentstyle[twoside,twocolumn,leqno,ltexpprt]{article}

\begin{document}

\newcommand{\op}[1]{{\mbox{\sc #1}}}
 \newcommand{\VEB}{{\sc veb}}
 \newcommand{\RAM}{{\sc ram}}
\newcommand{\vEB}{{van~Emde~Boas}}
\newcommand{\MSB}{{\mbox{{\sc MSB}}}}
\newcommand{\Div}{\mathbin{\bf div\/}}
\newcommand{\band}{\mathbin{\hbox{\bf bitwise-and\/}}}
\newcommand{\bor}{\mathbin{\hbox{\bf bitwise-or\/}}}
\newcommand{\bxor}{\mathbin{\hbox{\bf bitwise-xor\/}}}
\newcommand{\lshift}{\mathbin{\hbox{\bf left-shift\/}}}
\newcommand{\rshift}{\mathbin{\hbox{\bf right-shift\/}}}

\newcommand{\someskip}{\smallskip}
\newcommand{\Someskip}{\medskip}

\newcommand{\Paragraph}[1]{\medskip\noindent{\bf #1.\ }}
\newcommand{\Section}[1]{\section{#1.\ }}
\newcommand{\Subsection}[1]{\subsection{#1.\ }}

\title{\large\bf Chapter 21 \\
        \Large
Approximate Data Structures
with Applications
% \bigskip
\\
{\normalsize (Extended Abstract)}
% \bigskip
% \bigskip
}
\author{
{\it Yossi Matias}\ \thanks{\strut
AT\&T Bell Laboratories, 600 Mountain Avenue, Murray Hill, NJ 07974.
Email: matias@research.att.com.}
  \and
{\it Jeffrey Scott Vitter}\ \thanks{
Department of Computer Science,
Duke University, Box 90129,
% 205 North Building                         
Durham, N.C. 27708--0129.
Part of this research was done while the author was at Brown University.
This research was supported in part by
National   Science Foundation grant CCR--9007851 and by
Army Research Office grant~DAAL03--91--G--0035. Email: jsv@cs.duke.edu.}
  \and
{\it Neal E.~Young}\ \thanks{
Computer Science Department, Princeton University.
Part of this research was done while the author was at
UMIACS, University of Maryland, College Park, MD 20742
and was partially supported by NSF grants CCR--8906949 and CCR--9111348.
Email: ney@cs.princeton.edu.}
% \\
% \bigskip
% \bigskip
}

\newcommand{\old}[1]{}          % search for this to see old stuff

\date{}
%\date{\em Draft: not for distribution}
\maketitle

\pagestyle{myheadings}
\markboth{\sc Matias, Vitter, \& Young}
{\sc 
        Approximate Data Structures
 }

\begin{abstract}
In this paper we introduce the notion of {\em approximate data structures},
 in which a small amount of error is tolerated in the output.
Approximate data structures trade error of approximation for
 faster operation, leading to theoretical and practical speedups 
 for a wide variety of algorithms.
We give approximate variants of the \vEB{} data structure,
which support the same dynamic operations as the standard \vEB{}
data structure~\cite{VKZ-77,Mehlhorn:Book:I},
 except that answers to queries are approximate.
The variants support all operations in constant time
 provided the error of approximation is $1/{\mathop{\rm polylog}}(n)$,
 and in $O(\log\log n)$ time provided the error is $1/{\mathop{\rm
 polynomial}}(n)$, for $n$ elements in the data structure.

We consider the tolerance of prototypical algorithms 
 to approximate data structures.  We study in particular
Prim's minimum spanning tree algorithm,
 Dijkstra's single-source shortest paths algorithm,
 and an on-line variant of Graham's convex hull algorithm.
To obtain output which approximates the desired output 
 with the error of approximation tending to zero, 
 Prim's algorithm requires only linear time,
 Dijkstra's algorithm requires $O(m\log\log n)$ time,
 and the on-line variant of Graham's algorithm requires
 constant amortized time per operation.

\end{abstract}

\section{Introduction}
The \vEB{} data structure (\VEB{})~\cite{VKZ-77,Mehlhorn:Book:I} 
 represents an ordered multiset of integers.  
The data structure supports query operations for the current minimum 
 and maximum element, the predecessor and successor of a given element,
 and the element closest to a given number,
 as well as the operations of insertion and deletion.
Each operation requires $O(\log\log U)$ time,
where the elements are taken from a universe $\{0,...,U\}$.

\someskip
We give variants of the \VEB{} data structure
 that are faster than the original \VEB{},
 but only guarantee approximately correct answers.
The notion of approximation is the following:
 the operations are guaranteed to be consistent 
 with the behavior of the corresponding exact data structure
 that operates on the elements after they are mapped by 
 a fixed function~$f$.
For the multiplicatively approximate variant,
 the function~$f$ preserves the order of any two
 elements differing by at least a factor of some $1+\epsilon$.
For the additively approximate variant,
 the function~$f$ preserves the order of any two
 elements differing additively by at least some $\Delta$.

\someskip
Let the elements be taken from a universe $[1,U]$.
On an arithmetic \RAM{} with $b$-bit words,
    the times required per operation in our approximate
        data structures are as follows:
$$
% \begin{tabular}{|c|c|c|}  \hline
  \begin{tabular}{|c||c|c|}  \hline
   & multiplicative & additive \\
   & approx.~$(1+\epsilon)$ & approx.~$\Delta$ \\ 
%               \hline
                \hline \hline
time
\vrule height 17.5pt depth 12.5pt width0pt % hack to get more vertical space
 & $\displaystyle O\left( \log\log_b {\log U\over \epsilon}\right)$
     & $\displaystyle O\left( \log\log_b {U\over \Delta}\right)$ \\ \hline
\end{tabular}
$$

\Someskip
\noindent
Under the standard assumption
 that $b = \Omega(\log U + \log n)$,
 where $n$ is the measure of input size,
%Y the time required is 
   the time required is as follows:
%
% constant when $1/\epsilon$ and $U/\Delta$ are
% polylogarithmic, and it is $O(\log\log n)$ when $1/\epsilon$ and $U/\Delta$
% are polynomial.
%
$$
% \begin{tabular}{|c|c|c|}  \hline
\begin{tabular}{|c||c|c|}  \hline
                $\epsilon , \Delta/U $ 
   & 
%               $\mathop{\rm polylog}(nU)$ 
                $\mathop{1/}{\rm polylog}(nU)$ 
%               $(\log (nU))^{-\Omega(1)}$ 
   & 
%               $\mathop{\rm poly}(n)$ \\ \hline
                $\mathop{1/\exp(}{\rm polylog}(n))$ \\ \hline
                \hline \hline
time
\vrule height 17.5pt depth 12.5pt width0pt % hack to get more vertical space
 & 
        $O(1)$
 & 
        $ O(\log\log n) $ 
 \\ \hline
\end{tabular}
$$

\someskip
\Someskip
The space requirements of our data structures are $O(\log(U)/\epsilon)$
  and $O(U/\Delta)$, respectively.
The space can be reduced to close to linear in the number of
  elements by using dynamic hashing.
Specifically, the space needed is $O(|S| + |f(S)|\cdot t)$, where
 $S$ is the set of elements, $f$ is the fixed function mapping the
 elements of $S$ (hence, $|f(S)|$ is the number of distinct elements 
 under the mapping), and $t$ is the time required per operation.
The overhead incurred by using dynamic hashing is constant per memory
 access with high 
 probability~\cite{dietz-meyer,Dietzfelbinger:Gil:Matias:Pippenger:92}.
% Thus, when linear or nearly-linear space is sought, the complexity results 
%  should be interpreted as holding with high
%  probability; 
% they are otherwise deterministic, in which case the space used 
%  is $O(\log_{1+\epsilon}U)$ or $O(U/\Delta)$.
Thus, if the data structures are implemented to use nearly linear space,
 the times given per operation hold only with high probability.

%Y \Paragraph{Description of the data structure}
\Subsection{Description of the data structure}
\label{sec:description}
The approach is simple to explain,
 and we illustrate it for the multiplicative variant with 
     $\epsilon = 1$ and $b=1+\lfloor\log U\rfloor$.
Let $f(i) = \lfloor \log_2 i \rfloor$
 (the index of $i$'s most significant bit).
The mapping preserves the order of any two elements differing 
 by more than a factor of two
 and effectively reduces the universe size to $U' = 1+\lfloor\log U\rfloor$.
On an arithmetic \RAM{} with $b$-size words,
 a bit-vector for the mapped elements fits in a single word,
 so that successor and predecessor queries 
 can be computed with a few bitwise and arithmetic operations.
The only additional structures are a linked list of the elements 
% and a hash table mapping bit indices to list elements.
  and a dictionary mapping bit indices to list elements.

\someskip
In general, each of the approximate problems with universe size~$U$
 reduces to the exact problem with a smaller universe size~$U'$:
For the case of multiplicative approximation we have size 
%Y \marginpar{Y:Check this N:fixed }
\[
        U' = 2\log_2(U)/\epsilon
                = 
        O(\log_{1+\epsilon} U)\, ,
\] 
 and for the case of additive approximation 
\[ 
        U' = U/\Delta\, .
\]
Each reduction is effectively reversible, yielding an equivalence between
 each approximate problem and the exact problem with a smaller universe.
The equivalence holds generally for any numeric data type whose semantics 
 depend only on the ordering of the elements.
%Y \marginpar{Y:Back wards ? N:no}
The equivalence has an alternate interpretation:
 each approximate problem is equivalent to the exact problem 
 on a machine with larger words.
Thus, it precludes faster approximate variants
 that don't take advantage of fast operations on words.

\someskip
For universe sizes bigger than the number of bits in a word,
 we apply the recursive divide-and-conquer approach
 from the original \VEB{} data structure.
Each operation on a universe of size $U'$
 reduces to a single operation on a universe of size $\sqrt{U'}$
 plus a few constant time operations.
When the universe size is $b$, only a small constant number of 
 arithmetic and bitwise operations are required.
This gives a running time of $O(\log\log_b U')$, where $U'$ is 
 the effective universe size after applying the universe reduction 
 from the approximate to the exact problem.

\Subsection{Outline}
In the next section we motivate our development of approximate 
 \VEB{} data structures by demonstrating how they can be used in 
 three well-known algorithms: 
Prim's algorithm for minimum spanning trees, Dijkstra's shortest paths
 algorithm, and an on-line version of the Graham scan for finding convex 
 hulls.
Related work is discussed in Section~\ref{sec:Related}.
Our model of computation is defined in Section~\ref{sec:Model}.
In Section~\ref{sec:VEB}, we show how to construct our approximate 
 \VEB{} data structures and we analyze their characteristics.  
We make concluding remarks in Section~\ref{sec:Conclusions}. 

\section{Applications}
We consider three prototypical applications:
 to minimum spanning trees,
 to single-source shortest paths,
 and to semi-dynamic on-line convex hulls.
Our approximate minimum spanning tree algorithm 
 runs in linear time and is arguably simpler 
 and more practical than the two known linear-time MST algorithms.
Our approximate single-source shortest paths algorithm
 is faster than any known algorithm on sparse graphs.
Our on-line convex hull algorithm is also the fastest known in its class;
 previously known techniques require preprocessing and thus
 are not suitable for on-line or dynamic problems.
The first two applications are obtained by substituting
 our data structures into standard, well-known algorithms.
The third is obtained by a straightforward adaptation
 of an existing algorithm to the on-line case.
These examples are considered mainly as prototypical applications.
In general, approximate data structures
 can be used in place of any exact counterpart.

\someskip
Our results below assume a \RAM{} with a logarithmic word size 
 as our model of computation, described in more detail in 
 Section~\ref{sec:Model}.
The proofs are simple and are given in the full paper.

\subsection{Minimum spanning trees. }
For the minimum spanning tree problem, we show the following result about the
 performance of Prim's algorithm~\cite{jarnik:30,Prim:57,Dijkstra:59}
 when our approximate \VEB{} data structure is
 used to implement the priority queue:

\Someskip
\begin{theorem}
\label{thm:prim}
Given a graph with edge weights in $\{0,..,U\}$, Prim's algorithm,
 when implemented with our approximate \VEB{} 
 with multiplicative error $(1+\epsilon)$,
 finds a $(1+\epsilon)$-approximate minimum spanning tree
 in an $n$-node, $m$-edge graph in
 $O((n + m) \log(1 + (\log{1\over\epsilon})/\log\log nU))$ time.
\end{theorem}

\Someskip
% For $\epsilon = \Omega(\log^{-c}U)$, Theorem~\ref{thm:prim}
For $1/\epsilon \le \mathop{\rm polylog}(nU)$, Theorem~\ref{thm:prim}
 gives a linear-time algorithm.
This algorithm is arguably simpler and more practical
 than the two known linear-time MST algorithms.
This application is a prototypical example
 for which the use of an approximate data structure
 is equivalent to slightly perturbing the input.
Approximate data structures can be ``plugged in''
 to such algorithms without modifying the algorithm.

\subsection{Shortest paths. }
For the single-source shortest paths problem, we get the following
 result by using an approximate \VEB{} data structure as a priority queue 
 in Dijkstra's algorithm (see, e.g.,~\cite[Thm 7.6]{Tarjan:Book}):

\Someskip
\begin{theorem}
\label{thm:shortestpaths}
%Y \marginpar{N:reworded, fixed bug}
% Let $c$ be a constant, $c\ge 1$; let $\epsilon=\exp(-(\log^c n)\log\log U)$,
%  and let $\epsilon'=\exp(-(\log^c n)(\log\log U-1))$.
 Given a graph with edge weights in $\{0,...,U\}$
 and any $0 < \epsilon \le 2$, Dijkstra's algorithm,
 when implemented with our approximate \VEB{}
%with multiplicative error $(1+\epsilon')$,
 with multiplicative error $(1+\epsilon/(2n))$,
 computes single-source shortest path distances
%within a factor of $(1+\epsilon')$ in time $O(m\log\log n)$.
 within a factor of $(1+\epsilon)$ in 
 $O((n + m) \log (\log{n\over\epsilon}/\log\log U))$ time.
\end{theorem}

\Someskip
If $\log(1/\epsilon) \le \mathop{\rm polylog}(n) \log\log U$,
 the algorithm runs in $O((n + m)\log\log n)$ time ---
 faster than any known algorithm on sparse graphs ---
 and is simpler than theoretically competitive algorithms.
This is a prototypical example of an algorithm for which the error increases
 by the multiplicative factor at each step.
If such an algorithm runs in polynomial time,
 then $O(\log\log n)$ time per \VEB{} operation
 can be obtained with insignificant net error.
Again, this speed-up can be obtained 
 with no adaptation of the original algorithm.

\Paragraph{Analysis}
The proof of Theorem~\ref{thm:shortestpaths}
 follows the proof of the exact shortest paths algorithm
 (see, e.g.,~\cite[Thm 7.6]{Tarjan:Book}).
The crux of the proof is an inductive claim, saying that any vertex 
 $w$ that becomes labeled during or after the scanning of a vertex 
 $v$ also satisfies $dist(w)\ge dist(v)$, where $dist(w)$ is a
 so-called tentative distance from the source to $w$.
When using a $(1+\epsilon)$-approximate \VEB{} data structure to implement the
 priority queue, the inductive claim is
 replaced by \[
        dist(w)\ge dist(v)/(1+\epsilon/(2n))^i \, ,
\]      where vertex $v$ is
 the $i$th vertex to be scanned.
Thus, the accumulated multiplicative error is bounded by
\[
        (1+ \epsilon/(2n))^n 
        \le 
        e^{\epsilon/2}
        \le 
        (1+ \epsilon) \,.
\]
We leave the details to the full paper, and only note that it is not 
 difficult to devise an example where the error is actually accumulated
%exponentially at each iteration, to become $(1+\epsilon)^i$ after the
% $i$th iteration.
 exponentially at each iteration.

\subsection{On-line convex hull. }
Finally, we consider the semi-dynamic on-line convex hull problem.
In this problem, a set of planar points is processed in sequence.
After each point is processed, the convex hull of the points given so far
 must be computed. 
Queries of the form ``is $x$ in the current hull?''\ can 
 also be given at any time.
For the approximate version,
 the hull computed and the answers given must be consistent
 with a $(1+\Delta)$-approximate hull, which is
 contained within the true convex hull
 such that the distance of any point on the true hull
 to the approximate hull is $O(\Delta)$ times the diameter.

\someskip
We show the following result about the Graham scan algorithm \cite{Graham:72}
 when run using our approximate \VEB{} data structure:

\Someskip
\begin{theorem}\label{on-line CH}
The on-line $(1+\Delta)$-approx\-i\-mate convex hull can be computed
 by a Graham scan in constant amortized time per update
 if $\Delta\ge \log^{-c}n$
 for any fixed $c>0$, and in $O(\log\log n)$ amortized time per update
 if $\Delta\ge n^{-c}$.
\end{theorem}

\Someskip
This represents the first constant-amortized-time-per-query approximation 
 algorithm for the on-line problem.
This example demonstrates the usefulness of approximate data structures
 for dynamic/on-line problems.
Related approximate sorting techniques require preprocessing,
 which precludes their use for on-line problems.

\Paragraph{Analysis}
Graham's scan algorithm is based on scanning the points according 
 to an order determined by their polar representation, relative to 
 a point that is in the convex hull, and maintaining the convex hull
 via local corrections.
We adapt Graham's scan to obtain our on-line algorithm, as sketched below.
As an invariant, we have a set of points that are in the intermediate
 convex hull, stored in an approximate \VEB{} according to their
 angular coordinates.
The universe is $[0,2\pi]$ with a $\Delta$ additive error, which
 can be interpreted as the perturbation error of points in their angular
 coordinate, without changing their values in the distance coordinates.
This results in point displacements of at most $(1+\Delta)$ times 
 the diameter of the convex hull.

\someskip
Given a new point, its successor and predecessor in the \VEB{} are
 found, and the operations required to check the convex hull and, if 
 necessary, to correct it are carried on, as in Graham's 
 algorithm~\cite{Graham:72}.
These operations may include the insertion of the new point into the
 \VEB{} (if the point is on the convex hull) and the possible deletion
 of other points.
Since each point can only be deleted once from the convex hull, the
 amortized number of \VEB{} operations per point is constant.

\section{Related work}
\label{sec:Related}
Our work was inspired by and improves upon data structures
 developed for use in dynamic random variate generation
 by Matias, Vitter, and Ni~\cite{Matias:Vitter:Ni:93}.

\someskip
Approximation techniques such as rounding and bucketing have been 
 widely used in algorithm design.
This is the first work we know of that gives a general-purpose 
 approximate data structure.

%Y: \someskip
\Paragraph{Finite precision arithmetic} 
The sensitivity of algorithms to approximate data structures
 is related in spirit to the challenging problems that arise from
 various types of error in numeric computations.  
Such errors has been studied, for example, in the context of computational 
 geometry~\cite{%
Fortune:89,Fortune:Milenkovic:91,Greene:Yao:86,Guibas:Salesin:Stolfi:89,%
Milenkovic:88:PHD,Milenkovic:89,Milenkovic:89a}.
% Why is this relevant here ?
% The effect of perturbed vertex locations in the minimum spanning tree problem
% is similar to having imprecise inputs.
% Already said below above ?
% The effect of approximate data structures on the behavior of an algorithm
% is similar to the effect of non-accurate arithmetic in computations.
We discuss this further in Section~\ref{sec:Conclusions}.

%Y \someskip
\Paragraph{Approximate sorting}
Bern, Karloff, Raghavan, and Schieber~\cite{Bern:Karloff:Raghavan:Schieber:92}
 introduced {\em approximate sorting\/} and applied it
 to several geometric problems.
Their results include an $O((n\log\log n)/\epsilon)$-time algorithm
 that finds a $(1+\epsilon)$-approximate Euclidean minimum spanning tree.
They also gave an $O(n)$-time algorithm
 that finds a $(1+\Delta)$-approximate convex hull
 for any $\Delta \ge $ 1/polynomial.

\someskip
In a loose sense, approximate \VEB{} data structures 
 generalize approximate sorting.
The advantages of an approximate \VEB{} are the following.
An approximate \VEB{} bounds the error for each element individually.
Thus, an approximate \VEB{} is applicable
 for problems such as the general minimum spanning tree problem,
 for which the answer depends on only a subset of the elements.
The approximate sort of Bern {\em et al.~}bounds the {\em net} error,
 which is not sufficient for such problems.
More importantly, a \VEB{} is dynamic,
 so is applicable to dynamic problems such as on-line convex hull
 and in algorithms such as Dijkstra's algorithm 
 in which the elements to be ordered are not known in advance.
Sorting requires precomputation, so is not applicable to such problems.

% The main technical differences between approximate sorting
%  and approximate data structures are the following.
% Their approximate sort bounds the {\em net\/} error,
%  rather than bounding the error for each element {\em individually}.
% Thus, their technique does not apply directly
%  to problems such as the general minimum spanning tree problem
%  for which the answer depends on only a subset of the elements.
% More importantly, their method requires precomputation,
%  so it is inapplicable to dynamic or semi-dynamic problems such as the
%  on-line computation of convex hull, or for algorithms
%  such as Dijkstra's algorithm in which the elements to be ordered
%  are not known in advance.
% In general, however, the usefulness of approximate sorting
%  for geometric problems demonstrated by their work 
%  is an earlier evidence for the potential usefulness of approximate 
%  data structures.
 
%Y \Someskip
\Paragraph{Convex hull algorithms}
There are several relevant works for the on-line convex hull problem.
Shamos (see, e.g.,~\cite{PrepShamos}) gave an on-line algorithm for (exact)
 convex hull that takes $O(\log n)$ amortized time per update step.
Preparata~\cite{Preparata:79} 
 gave a {\em real-time\/} on-line (exact) convex hull algorithm
 with $O(\log n)$-time worst-case time per update step.
% Our on-line algorithm is based on Graham's scanning algorithm~\cite{Graham:72}.
Bentley, Faust, and Preparata~\cite{Bentley:Faust:Preparata:82}
 give an $O(n+1/\Delta)$-time algorithm
 that finds a $(1+\Delta)$-approximate convex hull.
Their result was superseded by the result of Bern {\em et~al.~}mentioned 
 above.
Janardan \cite{Janardan:91} gave an algorithm
 maintaining a fully dynamic ${(1+\Delta)}$-approximate convex hull
 (allowing deletion of points) in $O(\log(n)/\Delta)$ time per request.
Our on-line approximation algorithm is based on Graham's scan 
 algorithm~\cite{Graham:72} and can be viewed as a combination of 
 the algorithms by Shamos and by Bentley {\em et al.}, with the replacement 
 of an exact \VEB{} data structure by an approximate variant.
% Their algorithm (which is superseded by Bern et al's algorithm)
%  is similar to our modification of Graham's algorithm
% but does not dynamize as well.

%Y \someskip
\Paragraph{Computation with large words}
Kirkpatrick and Reich~\cite{Kirkpatrick:Reisch:84}
 considered exact sorting with large words,
 giving upper and lower bounds.
Their interest was theoretical,
 but Lemma~\ref{equivLemma}, which in some sense says that
 maintaining an approximate \VEB{} data structure
 is equivalent to maintaining an exact counterpart
 using larger words, suggests that lower bounds 
 on computations with large words are relevant to
 {\em approximate} sorting and data structures.

%Y \someskip
\Paragraph{Exploiting the power of RAM}
Fredman and Willard have considered a number of data structures 
 taking advantage of arithmetic and bitwise operations 
 on words of size $O(\log U)$.
In~\cite{Fredman:Willard:90}, 
 they presented the {\em fusion tree\/} data structure.
Briefly, fusion trees implement the \VEB{} data type
 in time $O(\log n/\log\log n)$.
They also presented an {\em atomic heap\/} data 
 structure~\cite{Fredman:Willard:FOCS:90}
 based on their fusion tree and used it to obtain 
 a linear-time minimum spanning tree algorithm
 and an $O(m+n\log n/\log\log n)$-time single-source shortest paths 
 algorithm.
Willard~\cite{Willard:92} also considered similar applications 
 to related geometric and searching problems.
Generally, these works assume a machine model similar to ours
   and demonstrate remarkable theoretical consequences of the model.
On the other hand, they are more complicated and involve larger constants.
% Our work can be viewed as exploring new avenues, that are interesting
% from both theoretical and practical points of view, in directions that
% were pointed out by the above works.

\someskip
Subsequent to our work Klein and Tarjan recently announced a randomized 
 minimum spanning tree algorithm that requires only expected linear 
 time~\cite{Klein:Tarjan:RMST}.
Arguably, our algorithm is simpler and more practical.

\section{Model of computation}
\label{sec:Model}
The model of computation assumed in this paper is a modernized version 
 of the {\em random access machine\/} (\RAM{}).  
Many \RAM{} models of a similar nature have been defined in the literature, 
 dating back to the early 1960s \cite{AHU:74}.  
Our \RAM{} model is a realistic variant of the logarithmic-cost 
 \RAM{}~\cite{AHU:74}:
the model assumes constant-time exact binary integer arithmetic
 ($+$, $-$, $\times$, $\Div$), 
 bitwise operations ($\lshift$, $\rshift$, $\bxor$, $\band$), 
 and addressing operations on words of size~$b$.  
Put another way, the word size of the \RAM{} is~$b$.
We assume that numbers are of the form $i+j/2^b$, 
 where $i$ and $j$ are integers with $0 \leq i,j < 2^b$,
 and that the numbers are represented with two words,
 the first holding $i$ and the second holding $j$.
For simplicity of exposition, we use the ``most-significant-bit'' 
 function $\MSB(x) = \lfloor \log_2 x \rfloor$; 
it can be implemented in small constant time
 via the previously mentioned operations
 and has lower circuit complexity than, e.g., division.

\section{Fast approximate data structures}
\label{sec:VEB}
This section gives the details of our approximate \VEB{} data structure.
First we give the relevant semantics and notations.
The operations supported are:
\\ 
\ \\
\indent $N \leftarrow \op{Insert}(x,d)$,\\
\indent $\op{Delete}(N)$,\\
\indent $N\leftarrow \op{Search}(x)$,\\
\indent $N\leftarrow \op{Minimum}(\ )$,\\
\indent $N\leftarrow \op{Maximum}(\ )$,\\
\indent $N\leftarrow \op{Predecessor}(N)$,\\
\indent $N\leftarrow \op{Successor}(N)$,\\
\indent $\phantom{N}\hbox to 0pt{\hss $d$} \leftarrow \op{Data}(N)$, and\\
\indent $\phantom{N}\hbox to 0pt{\hss $x$} \leftarrow \op{Element}(N)$.\\
\ \\
The $\op{Insert}$ operation and the query operations 
 return the {\em name\/} $N$ of the element in question.
The name is just a pointer into the data structure 
 allowing constant-time access to the element.
Subsequent operations on the element are passed this pointer
 so they can access the element in constant time.
$\op{Insert}$ takes an additional parameter $d$, 
 an arbitrary auxiliary data item.
$\op{Search}(x)$, where $x$ is a real number (but not necessarily an element), 
 returns the name of the largest element less than or equal to $x$.
%Y \marginpar{Y:Check this N:patched}
For the approximate variants, 
 the query operations are approximate in that the
 element returned by the query is within a $(1 + \epsilon)$
 relative factor or a $\Delta$ absolute amount of the correct value.
Operations $\op{Element}(N)$ and $\op{Data}(N)$,
 given an element's name~$N$, return
 the element and its data item, respectively.

\someskip
The universe (specified by $U$) and, for the approximate variants,
 the error of approximation ($\epsilon$ or $\Delta$)
 are specified when the data structure is instantiated.
% Moved to ``model of computation'' section:
% For the approximate variants, the elements may be real numbers.
% We assume that $\epsilon,\Delta \ge 1/U$.
% For definiteness, we assume that $U$ is a power of $2$,
% that the elements are of the form $i+j/U$, where $i$ and $j$ are integers
% with $1 \leq i,j \leq U$, 
% and that the elements are specified by two binary words holding $i$ and $j$.

\subsection{Equivalence of various approximations. }
\label{equivSec}

The lemma below assumes a logarithmic word-size \RAM{}.  
The notion of equivalence between data structures is that,
 given one of the data structures,
 the other can be simulated with constant-time overhead per operation.

\Someskip
\begin{lemma}
\label{equivLemma}
The problem of representing a multiplicatively
 $(1+\epsilon)$-approximate \VEB{} on universe $[1,U]$ is equivalent
 to the problem of representing an exact \VEB{} 
 on universe $\{0,1,...,O(\log_{1+\epsilon}U)\}$.

The problem of representing an additively
 $\Delta$-approximate \VEB{} on universe $[0,U]$ is equivalent
 to the problem of representing an exact \VEB{} 
%Y on universe $\{0,1,...,U/\Delta\}$.
   on universe $\{0,1,...,O(U/\Delta)\}$.
\end{lemma}
% The lemma assumes a logarithmic word-size \RAM{}.
% The notion of equivalence is that given one of the two data structures,
% the other can be simulated with constant-time overhead per operation.

\Someskip
\begin{proof}
Assume we have a data structure for the exact data type
 on the specified universe.
To simulate the multiplicatively approximate data structure,
 the natural mapping to apply to the elements (as discussed previously)
 is $x \mapsto \lfloor \log_{1+\epsilon}x \rfloor$.
Instead, we map~$x$ to approximately 
 ${1 \over \ln 2}(\log_{1+\epsilon} x ) 
 \approx (\log_2 x)/\epsilon$
and we use a mapping that is faster to compute:
%Y \marginpar{Y:Check this N:check again}
%Y Let $k = \lceil \log_2 {1\over \epsilon} \rceil$.  
%Y Let $x = i+j/2^b$.  
%Y Let $\ell = \MSB(i)$. 
Let $k = \lceil \log_2 {1\over \epsilon} \rceil$,  
 let $x = i+j/2^b$, and   
 let $\ell = \MSB(i)$. 
We use the mapping $f$ that maps $x$ to
\begin{eqnarray*}
% & & \ell \lshift k \\
% & & \mbox{} \bor (i \rshift (\ell - k)) \\
% & & \mbox{} \bxor (1 \lshift  k) \\
% & & \mbox{} \bor (j \rshift (b + \ell - k))
%Y & & \ell \lshift k \\
\nopagebreak
&\ell& \lshift (k) \\
&    & \mbox{} \bor (i \rshift (\ell - k)) \\
&    & \mbox{} \bxor (1 \lshift  k) \\
&    & \mbox{} \bor (j \rshift (b + \ell - k)) \, .
\end{eqnarray*}
% Recall that $\ell$ denotes the index of the most significant bit 
%  in the binary representation of $i$.
If $\ell < k$, then to right-shift by $(\ell-k)$
means to left-shift by $(k-\ell)$.
Note that in this case the fractional part of $x$ is shifted in.

\someskip
%Y \marginpar{Y:($k+1$)st ? N: think so}
This mapping effectively maps $x$ to the 
 lexicographically ordered pair $\langle \MSB(x), y\rangle$,
 where $y$ represents the bits with indices $(\ell -1)$ though 
 $(\ell -k)$ in $x$.
% the second through ($k+1$)st most significant bits of $x$.
The first part of the tuple distinguishes between any two $x$ values
 that differ in their most significant bit.
If two $x$ values have $\MSB(x) = \ell$,
 then it suffices to distinguish them if they differ additively by $2^{\ell-k}$.
The second part of the tuple suffices for this.

\someskip
Note that $f(1) = 0$ 
 and $f(U) < 2^{k+1} \log_2 U = O(\log_{1+\epsilon}U)$.
This shows one direction of the first part.
The other direction of the first part
 is easily shown by essentially inverting the above mapping,
 so that distinct elements map to elements that differ 
 by at least a factor of $1+\epsilon$.
Finally, the second part follows 
 by taking the mapping $(x \mapsto x \Div \Delta)$ and its inverse.
\end{proof}

\subsection{Implementations. }
Lemma~\ref{equivLemma} reduces the approximate problems
 to the exact problem with smaller universe size.
This section gives an appropriate solution to the exact problem.
If an approximate variant is to be implemented,
 we assume the elements have already been mapped by the constant-time 
 function $f$ in Lemma \ref{equivLemma}.
The model of computation is a \RAM{} with $b$-bit words.

\someskip
A {\em dictionary\/} data structure
 supports update operations $\op{Set}({\it key}, {\it value})$ and 
 $\op{Unset}({\it key})$ and query operation $\op{Look-Up}({\it key}) $
 (returning the value, if any, associated with the key).  
It is well known how to implement a dictionary
 by hashing in space proportional to the number of elements in the 
 dictionary or in an array of size proportional to the key space.
In either case, all dictionary operations require only constant time.
%Y (In the latter case, a well-known trick is required 
%Y  to instantiate the dictionary in constant time.)
%Y:
In the former case, the time is constant with high 
 probability~\cite{dietz-meyer,Dietzfelbinger:Gil:Matias:Pippenger:92};
 in the latter case, a well-known trick is required 
 to instantiate the dictionary in constant time.

\someskip
Each instance of our data structure will have a doubly-linked list of 
 element/datum pairs.
The list is ordered by the ordering induced by the elements.
The name of each element is a pointer to its record in this list.

\someskip
If the set to be stored is a multiset, as will generally be the case in 
 simulating an approximate variant,
 then the elements will be replaced by buckets, which are
 doubly-linked lists holding the multiple occurrences of an element. 
Each occurrence holds a pointer to its bucket.
In this case the name of each element is a pointer 
 to its record within its bucket.

\someskip
Each instance will also have a dictionary
 mapping each element in the set to its name.
If the set is a multiset, it will map each element to its bucket.
In general, the universe, determined when
 the data structure is instantiated, is of the form $\{L,...,U\}$.
Each instance records the appropriate~$L$ and $U$ values
 and subtracts $L$ from each element, 
 so that the effective universe is $\{0,...,U-L\}$.

\someskip
The ordered list and the dictionary suffice to support constant-time
 \op{Predecessor}, \op{Successor}, \op{Minimum}, and \op{Maximum} operations.
The other operations use the list and dictionary as follows.
$\op{Insert}(i)$ finds the predecessor-to-be of $i$
 by calling $\op{Search}(i)$, inserts $i$ into the list
 after the predecessor, and updates the dictionary.
If $S$ is a multiset, $i$ is inserted instead into its bucket
and the dictionary is updated only if the bucket didn't previously exist.
$\op{Delete}(N)$ deletes the element from the list (or from its bucket)
 and updates the dictionary appropriately.

\someskip
How $\op{Search}$ works depends on the size of the universe.
The remainder of this section describes $\op{Search}$ queries 
 and how $\op{Insert}$ and $\op{Delete}$ maintain
 the additional structure needed to support $\op{Search}$ queries.

\subsection{Bit-vectors. }
For a universe of size $b$,
 the additional structure required is a single $b$-bit word~$w$.
%Y As described in the Introduction, the word represents a 
As described in Section~\ref{sec:description}, the word represents a 
 bit vector; the $i$th bit is 1 iff the dictionary contains an element $i$.
$\op{Insert}$ sets this bit; 
$\op{Delete}$ unsets it if no occurrences of $i$ remain in the set.
Setting or unsetting bits can be done with a few constant time
 operations.

%Y To implement $\op{Search}(i)$, do the following.
The $\op{Search}(i)$ operation is implemented as follows.
If the list is empty or $i$ is less than the minimum element, return {\bf nil}.
Otherwise, let 
%Y \(
\[
        j \leftarrow \MSB(w \band ((1 \lshift i)-1)) \, ,
\]
%Y \)
%Y:
 i.e., let $j$ be
 the index of the most significant 1-bit in $w$
 that is at most as significant as the $i$th bit.
Return $j$'s name from the dictionary.

\Paragraph{Analysis}
On universes of size $b$,
 all operations require only a few constant-time operations.
If hashing is used to implement the dictionary,
 the total space (number of words) required at any time
 is proportional to the number of elements currently in the set.

% \subsection{Computing \MSB{} Quickly }
% If the function $\MSB(x) = \lfloor\log_2 x\rfloor$ is not available,
% the function $\LSB(x)$, defined to be the least significant 1-bit of $x$,
% will do just as well.
% It can be computed quickly as follows.
% First, note that $x \mapsto (x \bxor (x-1))$ maps $x$ to $2^{\lsb(x)+1}-1$.
% Next, note that there will be a small prime $p$ slightly larger than $b$
% such that $2^i \mod p$ is unique for $i \in \{0,...,p-1\}$,
% because $2$ generates the multiplicative mod-$p$ group.
% Thus, the function  $x \mapsto ((x \bxor {(x-1)}) \mod p)$
% maps $x$ to a number in $\{0,...,p-1\}$ that is uniquely
% determined by $\LSB(x)$.
% By precomputing an array $T$ of $p \approx b$ $b$-bit words
% such that $T[ (x \bxor (x-1)) \mod p] = \LSB(x)$,
% the \LSB{} function can subsequently be computed in a few operations.
% Examples of good $p$ include 
% 523 ($b={512}$), 269 ($b={256}$), 131 ($b={128}$), 67 ($b={64}$), 
% 37 ($b={32}$), 19 ($b={16}$), and 11 ($b={8}$).

%\subsection{Recursive Bitmaps for Larger Universes. }
\subsection{Intermediate data structure. }
The fully recursive data structure is a straightforward
 modification of the original \vEB{} data structure.
For those not familiar with the original data structure,
 we first give an intermediate data structure that is conceptually
 simpler as a stepping stone.
The additional data structures to support $\op{Search}(i)$ 
 for a universe $\{0,1,...,b^j-1\}$ are as follows.

\someskip
Divide the problem into $b+1$ subproblems:
 if the current set of elements is $S$,
 let $S_k$ denote the set $\{i\in S : i \Div b^{j-1} = k\}$.
Inductively maintain a \VEB{} data structure for each non-empty set $S_k$.
Note that the universe size for each $S_k$ is $b^{j-1}$.
Each $S_k$ can be a multiset only if $S$ is.

\someskip
Let $T$ denote the set $\{k : S_k \mbox{ not empty }\}$.
Inductively maintain a \VEB{} data structure for the set $T$.
The datum for each element $k$ is the data structure for $S_k$.
Note that the universe size for $T$ is $b$.
Note also that $T$ need not support multi-elements.

\someskip
Implement $\op{Search}(i)$ as follows.
If $i$ is in the dictionary, return $i$'s name.
Otherwise, determine~$k$ such that $i$ would be in $S_k$ if $i$ were in $S$.
Recursively search in $T$
 for the largest element $k'$ less than or equal to $k$.
If $k' < k$ or $i$ is less than the minimum element of $S_k$,
 return the maximum element of $S_{k'}$.
Otherwise, recursively search for the largest element 
 less than or equal to $i$ in $S_k$ and return it.

\someskip
$\op{Insert}$ and $\op{Delete}$ 
 maintain the additional data structures as follows.
$\op{Insert}(i)$ inserts $i$ recursively into the appropriate~$S_k$.
If $S_k$ was previously empty,
 it creates the data structure for $S_k$ and recursively inserts $k$ into $T$.
$\op{Delete}(N)$ recursively deletes the element from the appropriate~$S_k$.
If $S_k$ becomes empty, it deletes $k$ from $T$.

\Paragraph{Analysis}
Because the universe of the set $T$ is of size $b$,
 all operations maintaining $T$ take constant time.
Thus, each $\op{Search}$, $\op{Insert}$, and $\op{Delete}$
 for a set with universe of size $U = b^{j}$
 requires a few constant-time operations
 and possibly one recursive call on a universe of size $b^{j-1}$.
Thus, each such operation requires $O(j)=O(\log_b U)$ time.

\someskip
To analyze the space requirement, 
 note that the size of the data structure depends only
 on the elements in the current set.
Assuming hashing is used to implement the dictionaries,
 the space required is proportional to the number of elements
 in the current set plus the space that would
 have been required if the {\em distinct} elements of the current set
 had simply been inserted into the data structure.
The latter space would be at worst proportional to the time taken 
 for the insertions.
Thus, the total space required is proportional to the number of elements 
 plus $O(\log_b U)$ times the number of distinct elements.

\subsection{Full recursion. }
We exponentially decrease the above time
 by balancing the subdivision of the problem
 exactly as is done in the original \vEB{} data structure.

\someskip
The first modification 
 is to balance the universe sizes of the set $T$ and the sets $\{S_k\}$.
Assume the universe size is $b^{2^j}$.
Note that $b^{2^j} = b^{2^{j-1}}\times b^{2^{j-1}}$.
Define $S_k = \{i \in S : i \Div b^{2^{j-1}} = k\}$
 and define $T = \{k : S_k \mbox{ is not empty}\}$.
Note that the universe size of each $S_k$ {\em and} of $T$ is $b^{2^{j-1}}$.

\someskip
With this modification, $\op{Search}$,
 $\op{Insert}$, and $\op{Delete}$ are still well defined.
Inspection of $\op{Search}$ shows that
 if $\op{Search}$ finds $k$ in $T$, it does so in constant time,
 and otherwise it does not search recursively in $S_k$.
Thus, only one non-constant-time recursion is required,
 into a universe of size $b^{2^{j-1}}$.
Thus, $\op{Search}$ requires $O(j)$ time.

\someskip
$\op{Insert}$ and $\op{Delete}$, however,
 do not quite have this nice property.
In the event that $S_k$ was previously empty,
 $\op{Insert}$ descends recursively into both $S_k$ and $T$.
Similarly, when $S_k$ becomes empty,
 $\op{Delete}$ descends recursively into both $S_k$ and $T$.

\someskip
The following modification to the data structure fixes this problem,
 just as in the original \vEB{} data structure.
Note that $\op{Insert}$ only updates $T$ 
 when an element is inserted into an empty $S_k$.
Similarly, $\op{Delete}$ only updates $T$ 
 when the last element is deleted from the set~$S_k$.
Modify the data structure (and all recursive data structures)
 so that the recursive data structures exist only when $|S|\ge 2$.
When $|S|=1$, the single element is simply held in the list.
Thus, insertion into an empty set 
 and deletion from a set of one element
 require only constant time.
This insures that if $\op{Insert}$ or $\op{Delete}$ 
 spends more than constant time in $T$, 
 it will require only constant time in $S_k$.

\someskip
This modification requires that when $S$ has one element 
 and a new element is inserted,
 \op{Insert} instantiates the recursive data structures
 and inserts both elements appropriately.
The first element inserted will bring both $T$ and some $S_k$
 to size one; this requires constant time.
If the second element is inserted into the same set $S_k$ as the first element,
 $T$ is unchanged.  
Otherwise, the insertion into its newly created set $S_{k'}$ requires 
 only constant time.
In either case, only one non-constant-time recursion is required.

\someskip
Similarly, when $S$ has two elements and one of them is deleted,
 after the appropriate recursive deletions, 
 \op{Delete} destroys the recursive data structures 
 and leaves the data structure holding just the single remaining element. 
If the two elements were in the same set $S_k$, 
 then $T$ was already of size one, so only the deletion
 from $S_k$ requires more than constant time.
Otherwise, each set $S_k$ and $S_{k'}$ was already of size one,
 so only the deletion of the second element from $T$ 
 took more than constant time.

\Paragraph{Analysis}
With the two modifications, each $\op{Search}$, $\op{Insert}$,
 and $\op{Delete}$ for a universe of size $U = b^{2^j}$
 requires at most one non-constant-time recursive call,
 on a set with universe size $b^{2^{j-1}}$.
Thus, the time required for each operation is $O(j) = O(\log\log_b U)$.
As for the intermediate data structure,
 the total space is at worst proportional to the number of elements,
 plus the time per operation (now $O(\log\log_b U)$)
 times the number of distinct elements.

\section{Conclusions}
\label{sec:Conclusions}

The approximate data structures described in this paper
 are simple and efficient.
No large constants are hidden in the asymptotic notations---in fact, a
 ``back of the envelope'' calculation indicates significant speed-up
 in comparison to the standard \vEB{} data structure.
%Y How about adding here these calculations?
The degree of speed-up in practice will depend
 upon the machines on which they are implemented.
Machines on which binary arithmetic and bitwise operations on 
 words are nearly as fast as, say, comparison between 
 two words will obtain the most speed-up.
Practically, our results encourage the development of machines 
%Y \marginpar{Check PowerPC remark}
% like those based on the PowerPC chip 
% Perhaps better not to give only one example.
% In the journal we can collect a few example.
 which support fast binary arithmetic and bitwise operations
 on large words.
Theoretically, our results suggest the need for a model
% of computation that asymptotically distinguishes ``slow constant-time'' 
% from ``fast constant-time'' operations.
  of computation that more accurately measures the cost of operations
  that are considered to require constant time in traditional models.

% Refer to Floating POINT

\someskip
The applicability of approximate data structures to specific algorithms
 depends on the robustness of such algorithms to inaccurate 
 intermediate computations.
In this sense, the use of approximate data structures has an effect similar 
to computational errors that arise from the use of finite
 precision arithmetic.
In recent years there has been an increasing interest in studying 
 the effect of such errors on algorithms.
Of particular interest were algorithms in computational geometry.
Frameworks such as the ``epsilon geometry'' of
 Guibas, Salesin and Stolfi~\cite{Guibas:Salesin:Stolfi:89} 
 may be therefore relevant in our context.
The ``robust algorithms'' described by Fortune and
 Milenkovic~\cite{Fortune:89,Fortune:Milenkovic:91,%
 Milenkovic:88:PHD,Milenkovic:89,Milenkovic:89a}
% Sugihara:Iri:89} 
 are natural candidates for approximate data structures.

\someskip
Expanding the range of applications of approximate data structures
 is a fruitful area for further research.
Other possible candidates include
 algorithms in computational geometry that use the well-known sweeping 
 technique, provided that they are appropriately robust.
For instance, in the sweeping algorithm for the line arrangement problem 
 with approximate arithmetic, presented by 
Fortune and Milenkovic~\cite{Fortune:Milenkovic:91}, the priority
 queue can be replaced by an approximate priority queue with minor 
 adjustments, to obtain an output with similar accuracy.
If the sweeping algorithm of Chew and Fortune~\cite{Chew:Fortune:88}
 can be shown to be appropriately robust then the use of 
 the \vEB{} priority queue there can be replaced by an 
%approximate priority queue; an improved running time may imply better 
 approximate variant; an improved running time may imply better 
 performance for algorithms described 
 in~\cite{Bern:Karloff:Raghavan:Schieber:92}.

%TODO NEXT:
% Compute fast logarithm \& exponentiation approximations with
% applications~\cite{Johnston:92}
% 
% \section{[Practicality]}
% 
% Calculate when approx data structures are competitive.
% Is $2$-approximate competitive?  Will competitiveness
% increase as word size increases (Alpha chip)?
% 
% Floating point instead of integer inputs --- 
% nothing you can do theoretically unless range of exponents is restricted.
% In practice, range of exponents is small: have an exact chain of subproblems.

\bibliographystyle{plain}

%\bibliography{names,addr,full,approxDS,theory,theory1,tarjan,strings,confs}

% Insert this file in place of the bibliography lines at the end.

\end{document}